\documentclass[a4paper,11pt]{article}
\usepackage{pos}
\usepackage{braket}
\usepackage{amsmath}
\usepackage{graphicx} 
\usepackage{braket}
\usepackage{caption}
\usepackage{array}
\usepackage{booktabs}
\usepackage{ulem}
\usepackage{subcaption}
\usepackage{comment}
\usepackage{asymptote}
\usepackage{tikz,ifthen}

\makeatletter
\DeclareFontFamily{OMX}{MnSymbolE}{}
\DeclareSymbolFont{MnLargeSymbols}{OMX}{MnSymbolE}{m}{n}
\SetSymbolFont{MnLargeSymbols}{bold}{OMX}{MnSymbolE}{b}{n}
\DeclareFontShape{OMX}{MnSymbolE}{m}{n}{
    <-6>  MnSymbolE5
   <6-7>  MnSymbolE6
   <7-8>  MnSymbolE7
   <8-9>  MnSymbolE8
   <9-10> MnSymbolE9
  <10-12> MnSymbolE10
  <12->   MnSymbolE12
}{}
\DeclareFontShape{OMX}{MnSymbolE}{b}{n}{
    <-6>  MnSymbolE-Bold5
   <6-7>  MnSymbolE-Bold6
   <7-8>  MnSymbolE-Bold7
   <8-9>  MnSymbolE-Bold8
   <9-10> MnSymbolE-Bold9
  <10-12> MnSymbolE-Bold10
  <12->   MnSymbolE-Bold12
}{}

\let\llangle\@undefined
\let\rrangle\@undefined
\DeclareMathDelimiter{\llangle}{\mathopen}%
                     {MnLargeSymbols}{'164}{MnLargeSymbols}{'164}
\DeclareMathDelimiter{\rrangle}{\mathclose}%
                     {MnLargeSymbols}{'171}{MnLargeSymbols}{'171}
\makeatother

\title{Improvement in Autocorrelation Times Measured by the Master-Field Technique using Field Transformation HMC in 2+1 Domain Wall Fermion Simulations}
\ShortTitle{Improvement in Autocorrelation Times via the Master-Field Technique using FTHMC}

\author*[a]{Shuhei Yamamoto}
\author[a]{Peter Boyle}
\author[a]{Taku Izubuchi}
\author[b]{Luchang Jin}
\author[c]{Christoph Lehner}
\author[d]{Nobuyuki Matsumoto}

\affiliation[a]{Brookhaven National Laboratory,\\
  Upton, NY, USA}

\affiliation[b]{Department of Physics, University of Connecticut,\\
196A Auditorium Rd Unit 3046, Storrs, CT , USA}

\affiliation[c]{Fakult\"{a}t f\"{u}r Physik, Universit\"{a}t Regensburg, \\
Universit\"{a}tsstra$\beta$e 31, 93040 Regensburg, Germany}

\affiliation[d]{Department of Physics, University of Boston,\\
590 Commonwealth Ave 255, Boston, MA , USA}
\emailAdd{syamamoto@bnl.gov}

\abstract{The Field-Transformation Hybrid Monte-Carlo (FTHMC) algorithm potentially mitigates the issue of critical slowing down by combining the HMC with a field transformation, originally proposed by L\"{u}scher and motivated
as trivializing the theory.  For the transformation, we use a single invertible discrete smearing step inspired by the Wilson flow but which resembles a Jacobian-computable generalisation of the stout smearing step. This is applied to a system with Iwasaki gauge fields and 2+1 Domain-Wall fermions.  We have studied the effect of different smearing parameter values on autocorrelation times of Wilson-flowed energies with different flow time.  We have found a reduction of exponential autocorrelation times for infra-red observables
such as Wilson flowed energy densities and topological charge densities when a larger value of the smearing parameter is used.  The autocorrelation times of local observables are computed using an  approach akin to the master-field technique, allowing us to estimate the effect of the field transformation with different parameters based on a small number of configurations.}

\FullConference{The 41st International Symposium on Lattice Field Theory (LATTICE2024)\\
 28 July - 3 August 2024\\
Liverpool, UK\\}

\begin{document}
\maketitle

\section{Introduction}
In Ref.~\cite{Luscher:2009eq}, L\"{u}scher proposed an algorithm to mitigate the issue of critical slowing down using a transformation of gauge fields.  In this algorithm, a gauge field, $U$, is substituted by another field, $V$, using a diffeomorphism on a space of gauge fields as $U = \mathcal{F}_t(V)$
where $t$ is a parameter of the transformation.  Under this change of variable, 
\begin{equation*}
    Z = \int\mathcal{D}Ue^{-S(U)} =\int\mathcal{D}V\text{Det}[\mathcal{F}_*(V)]\,e^{-S(\mathcal{F}(V))} =\int\mathcal{D}V\,e^{-S_{FT}(V)}
\end{equation*}
so that we have a new action for the field $V$, namely
\begin{equation*}
    S_{FT} = S(\mathcal{F}_t(V)) - \text{ln Det }\mathcal{F}_{t*}(V).
\end{equation*}
In Ref.~\cite{Luscher:2009eq}, it is shown that there exists a perfect trivializing map $\mathcal{F}_t$ such that
\begin{equation*}
    \text{ln Det }\mathcal{F}_{t*}(V)\lvert_{t=1} = S(\mathcal{F}_t(V)) + \text{constant}.
\end{equation*}
For a perfect trivializing map, we have $S_{FT} = 0$ for $V$, and so $V$ is distributed uniformly.  
With a trivialized action, all modes of observables evolve at the same rate, reducing their autocorrelation times.
These ideal maps can be obtained as a solution to $\dot U_t = Z_t(U_t)U_t$
where $Z_t$ is a $\mathfrak{su}(3)$-field.

In this study, we use only a single discrete step as an approximation to the perfect trivializing map \cite{Foreman:2021ljl,Boyle:2022xor,Boyle:2024nlh} inspired by Wilson flow and stout smearing.  The numerical integration in Wilson flow
is replaced by a single the Euler step on each gauge link on the lattice:
\begin{equation}
    U(x,\mu) \to \mathcal{E}_{x,\mu}(y,\nu) = 
    \begin{cases}
        e^{Z_t(U)(x,\mu)}U(x,\mu) &\text{if } (y,\nu) = (x,\mu) \\
        U(y,\nu) &\text{otherwise.}
    \end{cases}
    \label{eq:wilson_evol}
\end{equation}
Here, 
\begin{align}
    Z_t(U_t)(x,\mu) &= \mathcal{P}(P_{\mu\nu}(x,\mu)) \equiv \frac{1}{2}(P_{\mu\nu}(x,\mu)-P_{\mu\nu}(x,\mu)^\dagger) - \frac{1}{6}\text{tr }[P_{\mu\nu}(x,\mu)-P_{\mu\nu}(x,\mu)^\dagger]\nonumber\\
    P_{\mu\nu}(x,\mu) &=\sum_{\nu \neq\pm\mu} \rho_{\mu,\nu}U(x,\nu)U(x+\hat{\nu},\mu)U(x+\hat{\mu},\nu)^\dagger U(x,\mu)^\dagger \label{eq:P}.
\end{align}
This gives a numerically cheap transformation which we might hope leads to a practical
gain in computer time.
The step size of the Euler step prescribed in Eq.~\ref{eq:wilson_evol} is absorbed into $\rho_{\mu,\nu} \equiv \rho$ in Eq.~\ref{eq:P}, and its effect on autocorrelation is investigated.  To make this technique suitable for computer simulation, a single link update in Eq.~\ref{eq:wilson_evol} is performed at even and odd sites and in each direction $\mu$ only once.  All links of a given polarisation on even or odd sites are updated in parallel.  Then, the overall link update consists of 8 independent updates and is more akin to familiar stout smearing.  

Compared to recent approaches seeking to completely trivialize the gauge theory and apply direct sampling schemes \cite{Kanwar:2024ujc}, this approach remains more similar to L\"{u}scher's HMC based algorithm  in structure while taking pains to minimize the cost of the field transformation even if the making
this transformation more approximate. 
The motivation is that rather than seeking to map fields from the trivial distribution in the strong coupling limit, we use the local averaging implicit in link smearing to make a wavelength-dependent transformation (in particular active in the ultraviolet) from our integration variables to our gauge field. Because the integration variable momenta are Gaussian distributed, this wavelength-dependent transformation can rescale the changes made to the physical gauge field differentially between the ultraviolet and infrared degrees of freedom, potentially changing the relative rate of evolution and avoiding large forces arising in the ultraviolet that might otherwise limit the infrared evolution
of the gauge field.

\section{Simulation Details}

This numerical experiment is conducted on a lattice of size $32^4$ with $\beta= 2.37$ and with $2+1$ dynamical Domain-Wall fermions of mass $m_l=0.0047, m_s=0.0186$.  We have performed several simulations with different HMC parameters, such as different $\rho$ values of 0.1, 0.112, 0.124, different gauge step sizes, $\delta\tau_G = 1/48, 1/96$, and different fermion step sizes $\delta\tau_F = 1/24,1/16,1/12,1/8$.  The nonzero value for $\rho$ indicates the use of field transformation in the HMC algorithm and will be referred to as FTHMC.  The algorithm with $\rho=0$ reduces to the usual HMC algorithm.  In the following, we focus on the runs with different $\rho$ and $\delta\tau_G$ but with fixed $\delta\tau_F = 1/24$ and trajectory length $\tau=1$.  Table \ref{tab:Stats_fixed_t_F} collects the total number of thermalized configurations from each run.  These simulations are carried out on Frontier and Andes at Oak Ridge National Laboratory. 
\begin{table}[hbt]
    \centering
    \begin{tabular}{|c|c|c|c|c|}
        \hline
        $\rho$ & 0.0 & 0.1 & 0.112 & 0.124\\
        \hline
$\delta\tau_G = 1/48$ & $233$ & $230$ & $188$ & $230$ \\
\hline
$\delta\tau_G = 1/96$ & $401$ & $232$  & $229$ & $229$ \\
\hline
$\delta\tau_G = 1/144$ & - & $230$ &-&-\\
\hline
    \end{tabular}\quad
    \caption{The number of trajectories for each ensemble after thermalization}
    \label{tab:Stats_fixed_t_F}
\end{table}

To see how reasonable the Monte Carlo (MC) chains are, the difference in the total simulation Hamiltonian between adjacent configurations in the chain, $dH$, has been computed.
The averages of $dH$ over MC chains are computed and summarized in Tab.~\ref{tab:dHs}.  
We have checked that the Creutz relation $\langle e^{-\Delta H}\rangle = 1$ is satisfied within statistics.
\begin{table}[hbt]
    \centering
    \begin{tabular}{|c|c|c|c|c|}
        \hline
        $\rho$ & 0.0 & 0.1 & 0.112 & 0.124\\
        \hline
$\delta\tau_G = 1/48$ & 0.026(6) & 0.006(6) & 0.009(5) & 0.03(1) \\
\hline
$\delta\tau_G = 1/96$ & 0.015(6) & $0 \pm 0.009$ & 0.009(8) & 0.017(7) \\
\hline
$\delta\tau_G = 1/144$ &  - & $0 \pm 0.008$ &-&-\\
\hline
    \end{tabular}
    \caption{$\braket{dH}$ for different runs based on configurations with Metropolis step.  $\delta\tau_F$ is fixed to $1/48$.}
    \label{tab:dHs}
\end{table}

\section{Measurements}

\begin{figure}[hbt]
\begin{minipage}{0.48\linewidth}
 \centering
    \includegraphics[width=0.8\textwidth]{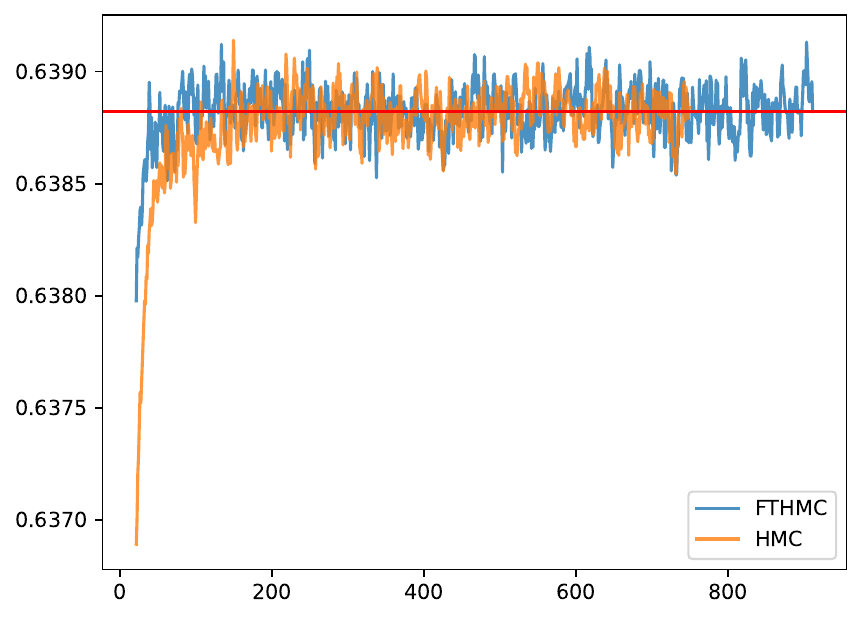}
    \caption{The plot of plaquette value vs. configuration index for HMC algorithm (orange line) and FTHMC algorithm (blue line).  The horizontal red line indicates the expected value of $0.6388238(37)$ for this ensemble \cite{RBC:2014ntl}.}
    \label{fig:plq}
\end{minipage}\hfill%
\begin{minipage}{0.48\linewidth}
   \centering
    \includegraphics[width=0.8\linewidth]{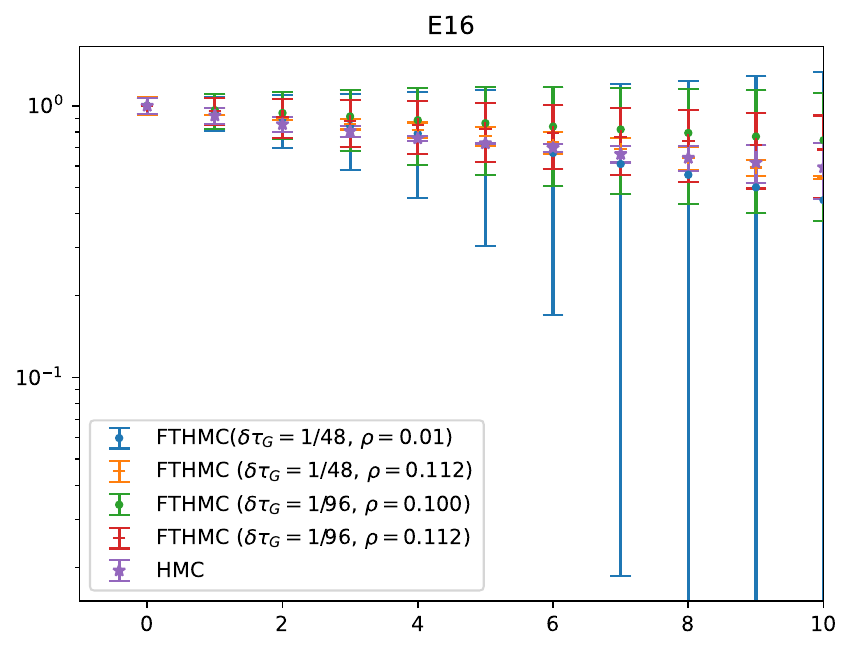}
    \caption{Autocorrelation coefficient (ACC) as a function of $t$ for Wilson-flowed energy E16, evaluated using the standard volume-averaged autocovariance $\Gamma^{(V)}(t)$.  This conventional estimator of ACC exhibits a large error. We will introduce an improved local autocorrelation approach.}
    \label{fig:exp_act_E16}
\end{minipage}
\end{figure}

As a first observable, we computed the plaquette.  Figure \ref{fig:plq} shows histories of plaquette values evaluated on each configuration from MC chains generated using the HMC and FTHMC algorithms.  Both chains begin with a hot configuration.  These MC chains are both observed to thermalize, as indicated by plaquette approaching to its expected value, which in this case is $0.6388238(37)$, taken from Ref.~\cite{RBC:2014ntl}.

\begin{figure}[htb]
    \centering
    \includegraphics[width=0.82\linewidth]{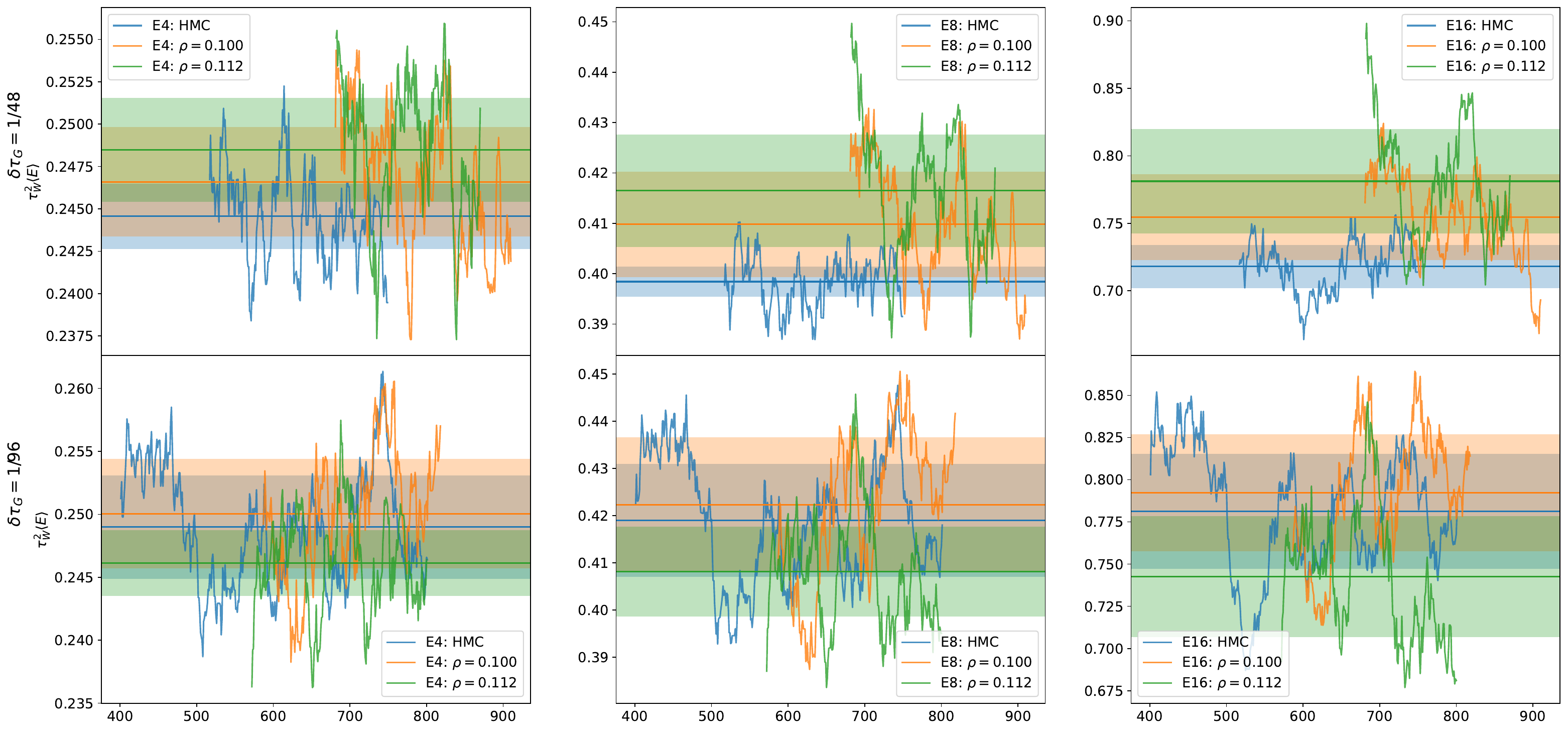}
    \caption{Comparison of Wilson flowed energy with different $\rho$ values for different flow time (column) and $\delta\tau_G=1/48,1/96$ (row).  The fluctuation is large so that estimation of ACC is difficult.}
    \label{fig:wflows_E_2}
\end{figure}

Another set of observables measured in this study is Wilson flowed energy with different flow time $\tau_W = 4,8,12,16$.  The flowed energy with longer flow time couples more strongly to slower mode and has a longer autocorrelation time, making it suitable for discerning different algorithms based on their effect on autocorrelation.  Histories of some flowed energy on thermalized configurations from different MC chains are shown in Fig.~\ref{fig:wflows_E_2}.  As can be seen from these figures, the value of $\tau_W^2 E(\tau_W)$ fluctuates wildly.  Determination of the central values as well as identification of repeated cycles are hard.  This makes it difficult to calculate their autocorrelation times.
\section{Autocorelation}

The autocorrelation of the flowed energy is considered in this study.  When we measure an observable $A(x)$ on each configuration in the MC chain, we obtain a sequence of measurements $\{a_i(x)\}_{i=1}^T$ where $T$ is the length of a finite Markov chain.  We write its volume average using double angle brackets as $\llangle A\rrangle = (1/V)\sum_x A(x)$
where $V$ is the lattice volume.  By averaging over statistically independent and infinitely many-times repeated simulations or, equivalently, infinite MC chain, the expectation value of $A(x)$ can be computed: $a \equiv \braket{a_i(x)} = \braket{\llangle A\rrangle}$.
We have introduced an abbreviation for the expectation value $a$.  In doing so, we have used wide-sense stationarity (WSS) for our HMC algorithms to eliminate the subscript $i$ for configurations in the chain.  The last equality follows from translational invariance.  For $\llangle A\rrangle$, we consider the volume autocovariance function, defined as $\Gamma^V(t) = \braket{\llangle a_i\rrangle\llangle a_{i+t}\rrangle}.$
Then, the autocorrelation coefficient is defined as the ratio $\rho^V(t) = \Gamma^V(t)/\Gamma^V(0)$.  As $T$ is finite here, we only estimate these expectation values.  The estimators are defined as appropriate averages over the finite MC chain.  For example, for $\Gamma^V(t)$, we use
\begin{equation*}
      \bar \Gamma^V(t) = \frac{1}{T-t}\sum_{i=1}^{T-t}(\llangle a_i\rrangle - \llangle\bar a\rrangle)(\llangle a_{i+t}\rrangle-\llangle\bar a\rrangle).
\end{equation*}
Figure \ref{fig:exp_act_E16} shows the estimated autocorrelation function based on the above formula.

The error on this plot is estimated using  Madras-Sokal Approximation \cite{Luscher:2004pav}:
\begin{equation*}
    \braket{\delta\bar\rho^{(V)}(t)^2} \simeq \frac{1}{N}\sum_{k=1}^{t+\Lambda}\left[\bar\rho^{(V)}(k+t)+\bar\rho^{(V)}(k-t) - 2\bar\rho^{(V)}(k)\bar\rho^{(V)}(t)\right].
\end{equation*}
Here, $\Lambda$ is set to 100, as $\Lambda\ge100$ gives a reasonable estimate of the error \cite{Luscher:2004pav}.

\subsection{Master-Field Technique}
The volume autocorrelation coefficient shown in Fig.~\ref{fig:exp_act_E16} suffers from a large error and does not reveal any information about autocorrelation of the volume average.  To obtain a clearer signal, a local observable $A(x)$ is considered.  More specifically, we considered the vacuum-subtracted local observable $A'(x) = A(x)-\llangle A\rrangle$ to reduce the bias of estimating $\Gamma_x(t) = \braket{(a_i(x) - a)(a_{i+t}-a)}$.  

Note $\mu = \braket{A'(x)} = a - a = 0$
so that
\begin{align*}
    \Gamma_x'(t) &= \braket{(a'_i(x) - \mu)(a_{i+t}'(x)-\mu)} = \braket{a'_i(x)a_{i+t}'(x)} \\
    &= \braket{(a_i(x)  - \llangle a_i\rrangle)(a_{i+t}(x)  - \llangle a_{i+t}\rrangle)} \equiv \braket{\mathcal{O}^i_t(x)}
\end{align*}   
is an unbiased estimator of autocovariance function of $A'(x)$.  Here, we introduced abbreviation $\mathcal{O}^i_t(x) \equiv (a_i(x)  - \llangle a_i\rrangle)(a_{i+t}(x)  - \llangle a_{i+t}\rrangle)$, treating it as a random variable over statistically independent MC chains.
As $\mathcal{O}^i_t(x)$ is an local observable, we can use stochastic locality to increase the statistics for $\Gamma_x'(t)$ and obtain the estimate with a smaller statistical error \cite{Luscher:2017cjh}.  The master-field estimator for  $\mathcal{O}^i_t(x)$ is
\begin{equation*}
    \llangle\mathcal{O}^i_t\rrangle \equiv \frac{1}{V}\sum_x \mathcal{O}^i_t(x).
\end{equation*}
The deviation of this estimator from the field-theoretic expectation value $O_t\equiv \braket{\mathcal{O}^i_t(x)}$ is of the order $\mathcal{O}(V^{-1/2})$ \cite{Luscher:2017cjh}.  We can further increase the statistics by summing over the index $i$:
\begin{equation*}
    \bar{\mathcal{O}}_t(x)\equiv\frac{1}{T-t}\sum_{i=1}^{T-t} \mathcal{O}^i_t(x).
\end{equation*}
The covariance of $\llangle\bar{\mathcal{O}}_t\rrangle$ with different simulation time separation is given by \cite{Luscher:2017cjh}
\begin{align*}
    \text{Cov}[\llangle\bar{\mathcal{O}}_s\rrangle,\llangle\bar{\mathcal{O}}_t\rrangle] 
    &\equiv  \braket{[\llangle\bar{\mathcal{O}}_s\rrangle-\braket{\mathcal{O}_s}][\llangle\bar{\mathcal{O}}_t\rrangle-\braket{\mathcal{O}_t}]} \\
    &= \frac{1}{V}\sum_y\braket{[\bar{\mathcal{O}}_s(y)-\braket{\mathcal{O}_s}][\bar{\mathcal{O}}_t(0)-\braket{\mathcal{O}_t}]} 
    \equiv \frac{1}{V}\sum_y C_{st}(y).
\end{align*}
Evaluation of this requires estimation of $C_{st}(y)$ for all spatial separation $y$, which is not feasible.  In exchange for a systematic error of $\mathcal{O}(e^{-R/\xi})$ with $\xi$ some characteristic length for correlation of $\bar{\mathcal{O}}_t$ at large distance, we can approximate the covariance \cite{Luscher:2017cjh} by truncating the sum over lattice as 
\begin{equation*}
    C_{st}(|y|\le R) \equiv \sum_{|y|\le R}C_{st}(y).
\end{equation*}
Also, we estimate each $C_{st}(x)$ by $\llangle C_{st}(x)\rrangle$.  As $\llangle\bar{\mathcal{O}}_t(x)\rrangle$ is an estimator for $\Gamma_x'(t)$, this estimate of the covariance can then be used in estimating the error of $\rho(t)$ by inserting them into
\begin{equation*}
    \text{Var}[\rho(t))] = (\rho(t))^2\left(\frac{\text{Var}[\llangle\bar{\mathcal{O}}_t(x)\rrangle]}{\llangle\bar{\mathcal{O}}_t(x)\rrangle^2}+\frac{\text{Var}[\llangle\bar{\mathcal{O}}_0(x)\rrangle]}{\llangle\bar{\mathcal{O}}_0(x)\rrangle^2}-2\frac{\text{Cov}[\llangle\bar{\mathcal{O}}_t(x)\rrangle,\llangle\bar{\mathcal{O}}_0(x)\rrangle]}{\llangle\bar{\mathcal{O}}_t(x)\rrangle\llangle\bar{\mathcal{O}}_0(x)\rrangle}
   \right).
\end{equation*}

\begin{figure}[htb]
    \centering
    \includegraphics[width=0.8\linewidth]{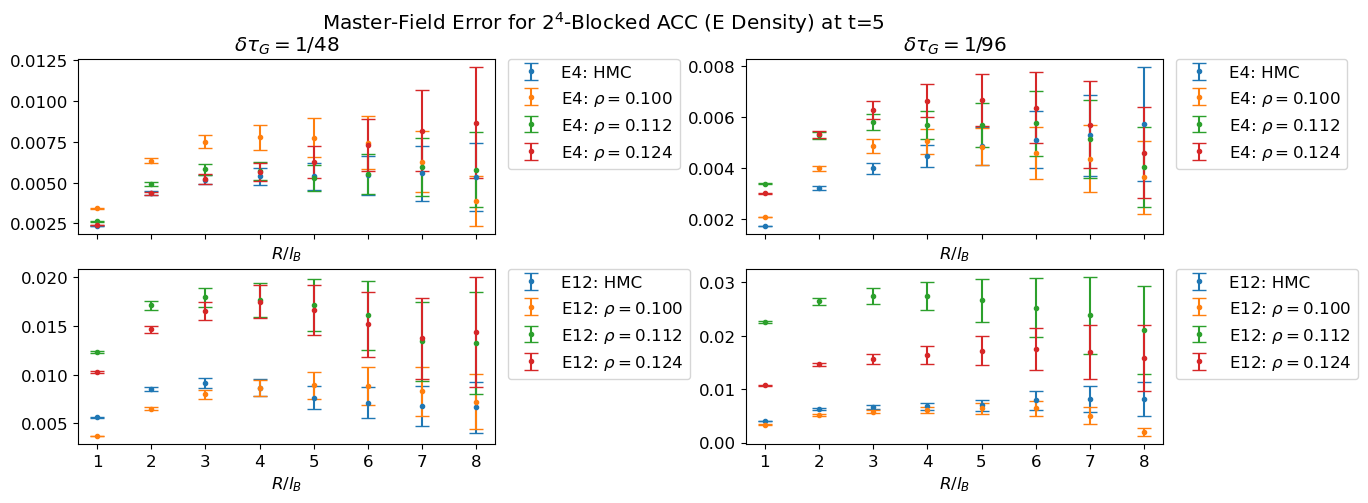}
    \caption{The plots of estimated $\sigma_{\rho(t)}$ as a function of $R/2$ for flowed energy with different flow time, of 4 and 16 shown as examples, and gauge step size $\delta\tau_G$.  The plots are observed to plateau around $R/2 = 4$.}
    \label{fig:MF_error}
\end{figure}

We expect $C_{st}(|y|\le R)$, and thus estimated $\sigma_{\rho(t)}$, to saturate as $R$ increases due to exponential suppression of spatial correlation.  To reduce computational intensity, the blocking of lattice into $2^4$ blocks is introduced, following the suggestion in Ref.~\cite{Bruno:2023vhs}.  The plots of estimated $\sigma_{\rho(t)}$ as a function of $R/2$ for flowed energy with different flow time and gauge step size $\delta\tau_G$ is shown in Fig~\ref{fig:MF_error}.
The error on the plots are computed using the formulae for the covariance of covariances presented in Ref.~\cite{Bruno:2023vhs}.  As can be seen from the graph, the error saturates at $R/2 = 4$.  Using this value of $R$, we computed the estimate of local autocorrelation of flowed energy.  

\begin{figure}[ht]
        \begin{minipage}[b]{0.5\linewidth}
            \centering
            \includegraphics[width=\textwidth]{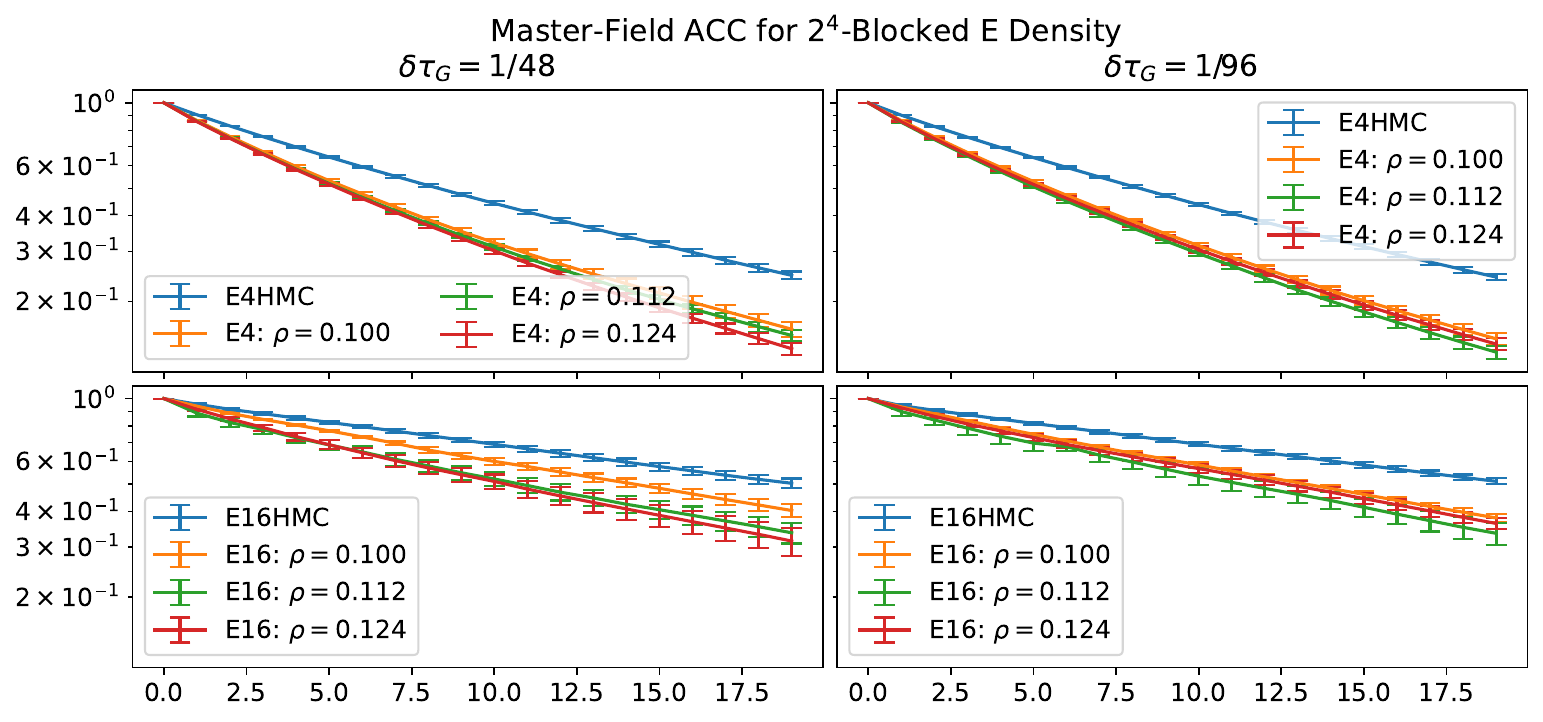}
            \label{fig:a}
        \end{minipage}
        \hspace{0.1cm}
        \begin{minipage}[b]{0.5\linewidth}
            \centering
            \includegraphics[width=\textwidth]{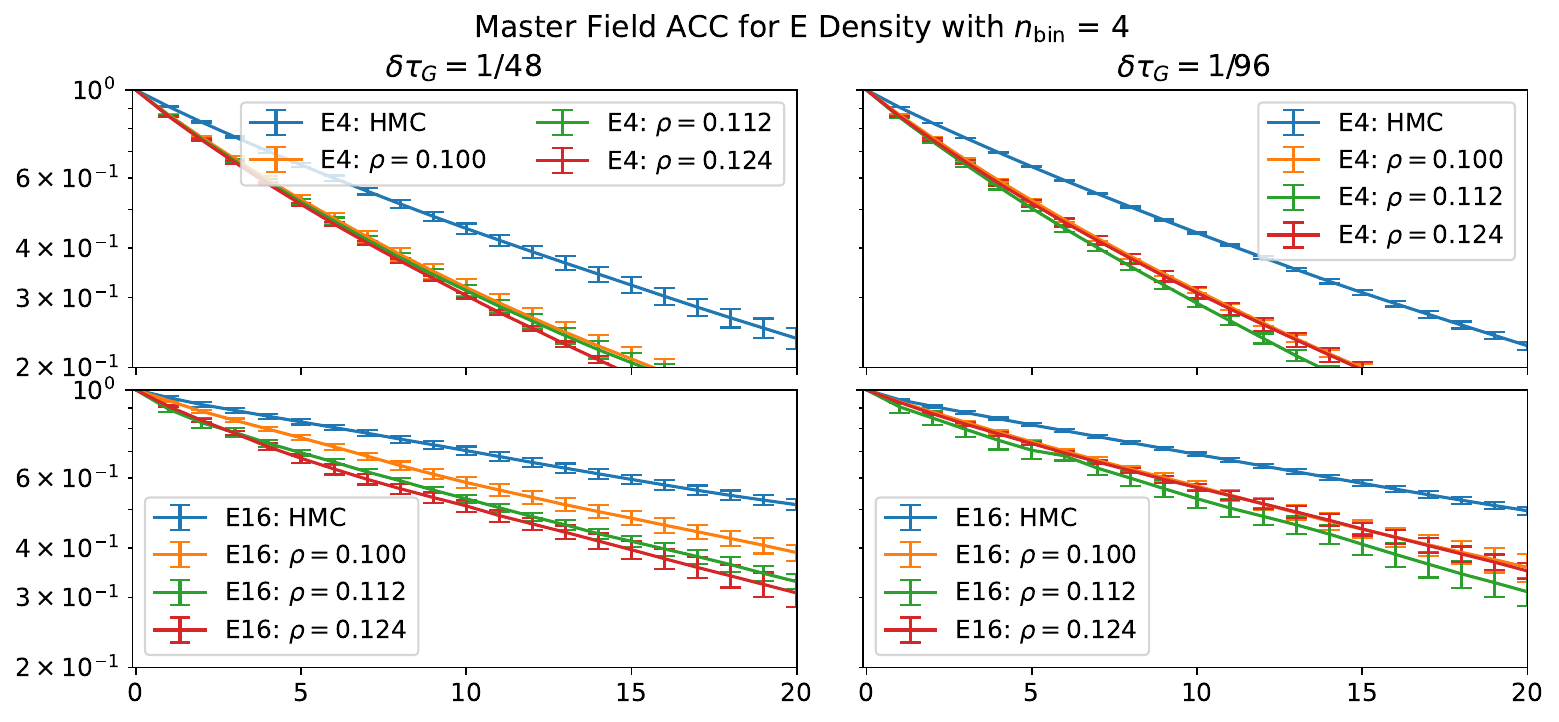}
            \label{fig:b}
        \end{minipage}
        \caption{Autocorrelation coefficient (ACC) estimated using master-field technique for $\tau_w=4,16$.  The error of ACC's in the left plot is estimated also using master-field technique.  The error in the right by binning of bin size of 4. ACC with FTHMC is observed to diminish faster than with HMC.  Also, two different estimation methods produce similar plots.}
        \label{fig:ACC_compare}
    \end{figure}

On the other hand, we can also make a quick estimate of local autocorrelation coefficient using a binning method.  For this, we first divide the MC chain into $n_{\rm bin}$ bins, estimate $\rho(t)\lvert_{b}$ using master-field technique on each bin, $b$, and take the average and standard deviation of the mean as the estimate of $\rho(t)$ and its error, respectively.  As long as each bin is long enough as compared to autocorrelation time of interest and as there are enough many bins to allow us to use Central Limit Theorem, estimation of error on autocorrelation function via binning is valid.  The comparison of estimated autocorrelation coefficients (ACC) based on master-field technique and binning is shown in Fig.~\ref{fig:ACC_compare}.  They are not significantly different albeit the small number of bins used so that we can use binning for quick estimates of autocorrelation time, as it involves a less amount of computation.  Note that we now have much smaller noise and clearer signal, as compared to volume-averaged autocorrelation coefficients.
\vspace{-0.5cm}
\section{Conclusion and outlook}
The master-field autocorrelation method is intended to be used for parameter optimization, and plots of ACC presented in Fig.~\ref{fig:ACC_compare} allows us to discern effects of different algorithms and parameters on autocorrelation based on a small number of configurations, serving our purpose.  However, to gain an idea on effects of different algorithms with various parameter choices, exponential autocorrelation times, $\tau_{\rm exp}$, for each flowed energy with different HMC parameters are estimated by fitting $e^{-t/\tau}$ to the data.  Due to a limited number of configurations, the attempt to estimate the error was not made. Table \ref{tab:exp_fit_summary} collects $\tau_{\rm exp}$ for each algorithm and HMC parameter choice, and Tab.~\ref{tab:exp_fit_summary_ratio} the ratios of $\tau_{\rm exp}(\rho=0.0,\delta\tau_G=1/48)$ for HMC to $\tau_{\rm exp}$ with other HMC parameters.  The table suggests that FTHMC reduces autocorrelation time by a factor of about $1.5$.  At physical pion mass, we expect to see full improvement of HMC simulation by a factor of around 1.5.

In the period between the presentation of this conference talk and our proceedings, we have made substantive additional progress on several fronts. The software implementation in \texttt{Grid} \cite{Boyle:2015tjk} for calculating $\text{ln Det }\mathcal{F}_{t*}(V)$ and force from this action has been carefully optimized and the FTHMC overhead (link smearing, force chain rule, and Jacobian force) now run around four times faster. 
These optimizations took the form of three types. Firstly, the data parallel layer was initially performing redundant work on both checkerboards of the lattice when only gauge links on a single checkerboard were required to be updated. Secondly, latencies in GPU kernel launch were substantial, and greater GPU utilization was obtained by fusing multiple loops into a single kernel call. Finally, the staple assembly was accelerated
by using a single halo-exchange and generalized stencil objects in place of repeated covariant shifts.  
The overhead now occupies 2.2\% of the total run time when we start from a thermalized configuration using a single node with 8 GPU's on Frontier, making our conclusion even stronger.

The FTHMC approach has also been independently implemented and optimized in the Grid Python Toolkit (GPT) framework \cite{GPT}, serving as a further check on the methods.  In a further study within GPT, the algorithm was tuned and combined with longer trajectories up to $\tau=8$.
Both FTHMC and usage of longer trajectories proved to accelerate decorrelation of infrared observables studied here.  They also improved global topological charge sampling at physical quark masses in large-volume and fine-lattice simulations. Gains from both effects are additive, and we are studying in more detail these promising algorithmic directions which may lead to deeper understanding and further gains in the future. The method is now being used in production of a new RBC-UKQCD 2+1f ensemble with physical quark masses, $128^3\times 288$ volume, and $a^{-1} = 3.5$ GeV.  An overall gain in autocorrelation
is around 3.5. The preservation of topological tunneling is demonstrated and will enable us to take a continuum limit based on four lattice spacings for a number of physical quantities.
The method will be the subject of an in-preparation journal publication.
\section{Acknowledgements}

SY and PB have been supported in part by the Scientific Discovery
through Advanced Computing (SciDAC) program LAB 22-2580. PB has been supported by US DOE Contract DESC0012704(BNL). We have made use of computer allocated under the Innovative and Novel Computational Impact on Theory and Experiment (INCITE) program on the Frontier and Andes supercomputers at ORNL.   We
acknowledge the EuroHPC Joint Undertaking for awarding this project access to the EuroHPC supercomputer
LUMI, hosted by CSC (Finland) and the LUMI consortium through a EuroHPC Extreme Scale Access call.

\begin{table}[bht]
    \centering
    \begin{tabular}{|c|c|c|c|c|c|c|c|c|}
    \toprule
     {}&  \multicolumn{4}{|c|}{$\delta\tau_G=1/48$} & \multicolumn{4}{|c|}{$\delta\tau_G=1/96$}\\
    \midrule
      $\rho$  & $\tau_W = 4$ & $\tau_W = 8$ & $\tau_W = 12$ & $\tau_W = 16$ & $\tau_W = 4$ & $\tau_W = 8$ & $\tau_W = 12$ & $\tau_W = 16$ \\
      \hline
0.0 & 14.28 & 22.439 & 26.6961 & 27.822 & 14.0 & 22.768 & 27.613 & 29.6098 \\
\hline
0.100 & 11.2 & 17.3 & 20.628 & 21.68 & 10.53 & 15.776 & 19.071 & 20.683 \\
\hline
0.112 & 10.87 & 15.62 & 17.97 & 19.2 & 10.0 & 14.862 & 17.25 & 18.86 \\
\hline
0.124 & 10.14 & 14.8 & 16.96 & 17.68 & 10.4 & 15.686 & 18.884 & 20.453 \\
\bottomrule
    \end{tabular}
    \caption{Estimates of exponential autocorrelation times $\tau_{\rm exp}$ computed by fitting $e^{-t/\tau_{\rm exp}}$ to the ACC for different Wilson flow time $\tau_W$, $\rho$, and $\delta\tau_G$ values}
    \label{tab:exp_fit_summary}
\end{table}

\begin{table}[bht]
    \centering
    \begin{tabular}{|c|c|c|c|c|c|c|c|c|}
    \toprule
     {}&  \multicolumn{4}{|c|}{$\delta\tau_G=1/48$} & \multicolumn{4}{|c|}{$\delta\tau_G=1/96$}\\
    \midrule
      $\rho$  & $\tau_W = 4$ & $\tau_W = 8$ & $\tau_W = 12$ & $\tau_W = 16$ & $\tau_W = 4$ & $\tau_W = 8$ & $\tau_W = 12$ & $\tau_W = 16$ \\
      \hline
0.0 & 1&1&1&1& 1.019 & 0.9855 & 0.96681 & 0.93961 \\
\hline
0.100 & 1.275 & 1.2967 & 1.2942 & 1.2832 & 1.356 & 1.4224 & 1.3998 & 1.3451 \\
\hline
0.112 & 1.313 & 1.436 & 1.485 & 1.4487 & 1.428 & 1.5098 & 1.5476 & 1.4754 \\
\hline
0.124 & 1.408 & 1.516 & 1.574 & 1.5736 & 1.373 & 1.4305 & 1.4137 & 1.3603 \\
\bottomrule
    \end{tabular}
    \caption{The ratios of $\tau_{\rm exp}(\rho=0.0,\delta\tau_G=1/48)$ for HMC to $\tau_{\rm exp}$ with other HMC parameters}
    \label{tab:exp_fit_summary_ratio}
\end{table}

\bibliography{lattice2024}


\end{document}